\documentclass[onecolumn,showpacs,draft]{revtex4}
\usepackage{epsfig}
\usepackage{colordvi}
\usepackage{bm}
\usepackage{amsmath}
\usepackage{amssymb}
\usepackage{graphicx}

\begin{document}

\title{Nonlocal Gate Of Quantum Network Via Cavity Quantum Electrodynamics}
\author{Xiang-Fa Zhou}
\email{xfzhou@mail.ustc.edu.cn}
\author{Yong-Sheng Zhang}
\email{yshzhang@ustc.edu.cn}
\author{Guang-Can Guo}
\email{gcguo@ustc.edu.cn}
\affiliation{\textit{Laboratory of
Quantum Information, University of Science and Technology of
China, Hefei, Anhui 230026, People's Republic of China}}

\begin{abstract}
We propose an experimentally feasible scheme to realize the nonlocal gate
between two different quantum network nodes. With an entangled qubit (ebit)
acts as a quantum channel, our scheme is resistive to actual environment
noise and can get high fidelity in current cavity quantum electrodynamics
(C-QED) system.
\end{abstract}

\maketitle

Quantum information science aims to develop new theories and methods for
processing information at the quantum level. Superposition and entanglement
play the central roles in this field and make it quite different from
classical case. Many interesting schemes have been proposed during the past
ten years \cite{dense,tele,cryp}. Most of these schemes rely on the
realization of Quantum networks \cite{cirac}, which are considered as
spatially separated nodes connected by quantum communication channels.
Actually quantum communication and operation between different nodes make it
possible for various of novel applications, such as distributed quantum
computation \cite{cirac2}, quantum key distribution \cite{keydistribution},
etc. As to physical implementation, the network nodes are often thought as
clusters of trapped atoms and ions, which are the ideal candidates for their
long-lived internal states. On the other hand, photon represents the best
carriers for fast and reliable communication over long distance. Thus how to
implement non-local interaction between spatially distant quantum network
nodes seems an interesting question.

Earlier schemes for the realization of the interaction between two different
network nodes \cite{cirac} rely on the transmission of a single photon
through quantum channel. Thus environment noise is unavoidably introduced.
Moreover, the transmitted signals can not be amplified in quantum area
because of the so-called non-cloning theorem \cite{wooters}. When photon
absorption or transmission error occurs, the probability of success and the
fidelity will be dramatically decreased. However, if entanglement is
concerned here, thing will be changed. In fact, entanglement purification
\cite{puri} together with quantum repeaters \cite{repeater} provides an
efficient method against the environment noise.

In this paper, we consider the most simple case for future's network quantum
computation. With the help of high-quality cavities, we use a pair of
entangled photons to realize the Controlled-NOT (C-NOT) operation between
two spatially-separated nodes. Our scheme is probabilistic, insensitive to
actual environment noise and can reach high fidelity with finite efficient
detectors.

The quantum network node consists of a trapped-atom inside a high-Q cavity
in our scheme. The atom has a standard \textquotedblleft $\Lambda $%
\textquotedblright\ type internal states, and the relevant level diagram is
shown in Fig. 1. This kind of level structure can be found in alkali atom
with $|0\rangle $ and $|1\rangle $ denoting hyperfine manifolds of the
ground state. A cavity mode $a_{h}$ resonantly couples the atom transition $%
|0\rangle \longrightarrow |e\rangle $ with horizontal ($h$) polarization.
The ebits concerned here are to be polarization-entangled photons shining
onto the cavity through a polarizing beam splitter (PBS) (Fig. 3). Only the $%
h$ component can transmit the PBS and resonantly drives the corresponding
cavity mode $a_{h}$. The vertical ($v$) component is reflected by the PBS
and the mirror. This kind of C-QED system has been proposed by Duan \textit{%
et al.} \cite{Duan} to realize the interaction between two single-photon
pulses. Here with the help of a pair of entangled photons we use similar
arrangement to implement the C-NOT operation between two distant network
nodes.

Before describing the detailed model, we want to summarize the
basic ideas of our scheme first. We consider two network nodes $A$
and $B$ with $A$ as
the control part and $B$ as the target, and set the initial states to be $%
|\varphi _{A}\rangle =\alpha |0\rangle _{A}+\beta |1\rangle _{A},\ |\varphi
_{B}\rangle =a|0\rangle _{B}+b|1\rangle _{B}$. Generally, we assume that
Alice holds system $A$ and Bob holds system $B$. To implement a quantum
C-NOT gate, besides one bit of classical communication in each direction,
one shared ebit is also necessary \cite{nonlocal gate}. Here we choose the
ebit to be a polarization-entangled photon pair $|\varphi \rangle
_{A_{1}B_{1}}=\frac{1}{\sqrt{2}}(|h\rangle _{A_{1}}|v\rangle
_{B_{1}}+|v\rangle _{A_{1}}|h\rangle _{B_{1}})$. The whole scheme can be
understood from the following steps.

Step (A): First, the incoming photon $A_{1}$ passes a local Hadamard gate
and then enters into the C-QED system to realize the CPF operation between
the atom and the photon. The pulse is then reflected and undergoes another
Hadamard rotation. This process actually realizes the C-NOT operation with
the node $A$ as the control qubit. After this process the combined state of $%
A$, $A_{1}$ and $B_{1}$ becomes
\begin{equation}
\frac{1}{\sqrt{2}}[\alpha |0\rangle _{A}(\left\vert 1\right\rangle
_{A_{1}}\left\vert 0\right\rangle _{B_{1}}+|0\rangle _{A_{1}}|1\rangle
_{B_{1}})+\beta |1\rangle _{A}(|0\rangle _{A_{1}}|0\rangle
_{B_{1}}+|1\rangle _{A_{1}}|1\rangle _{B_{1}})]\text{,}
\end{equation}%
where we have set $|h\rangle \longrightarrow |1\rangle $ and $|v\rangle
\longrightarrow |0\rangle $. The photon now reaches the PBS and causes a
click on one of the two detectors $D_{A_{v}}$ and $D_{A_{h}}$. Alice then
registers the corresponding result $r_{a}$.

Step (B): Similar to step (A), after a C-NOT operation between $B$ and $%
B_{1} $ with the photon $B_{1}$ as the control qubit, Bob performs another
Hadamard operation on part $B_{1}.$ For example, if $r_{a}=D_{A_{v}}$, after
this step we can get the combined state of $A$, $B$ and $B_{1}$ as%
\begin{eqnarray}
&&\left\vert 1\right\rangle _{B_{1}}(-\alpha a\left\vert 01\right\rangle
-\alpha b\left\vert 00\right\rangle +\beta a\left\vert 10\right\rangle
+\beta b\left\vert 11\right\rangle )_{AB}  \notag \\
&&+\left\vert 0\right\rangle _{B_{1}}(\alpha a\left\vert 01\right\rangle
+\alpha b\left\vert 00\right\rangle +\beta a\left\vert 10\right\rangle
+\beta b\left\vert 11\right\rangle )_{AB}.
\end{eqnarray}%
Bob then detects the reflected photon on $D_{B_{v}}$ and $D_{B_{h}}$ and
registers the result $r_{b}$.

Step(C): Now Alice and Bob commute their measurement results
through classical channel. Proper local rotations (Table. 1) are
then performed on nodes $A$ and $B$ respectively according to the
information they have got. For example, if $r_{a}=D_{A_{v}}$, and
$r_{b}=D_{B_{h}}$, after local operations $\sigma _{z_{A}}$ and
$\sigma _{x_{B}}$, they can get the desired state $\alpha
a|0\rangle _{A}|0\rangle _{B}+\alpha b|0\rangle _{A}|1\rangle
_{B}+\beta a|1\rangle _{A}|1\rangle _{B}+\beta b|1\rangle
_{A}|0\rangle _{B}$.

In the ideal case, the detectors always register a click. However,
perfect single-photon detectors are unreachable with current
technology. When either one of Alice or Bob doesn't get a click,
they have to reset the whole system and repeat these steps.\ Fig.
2 depicts a general quantum circuit to perform the non-local gate,
where we have decomposed the C-NOT gate into a controlled-phase
gate plus two Hadamard rotations. A possible experimental setup
can be found in Fig. 3.

\qquad \qquad \qquad \qquad \qquad \qquad \qquad \qquad \qquad
Table. 1

\qquad \qquad \qquad \qquad \qquad \qquad \qquad \qquad \qquad Fig. 1

\qquad \qquad \qquad \qquad \qquad \qquad \qquad \qquad \qquad Fig. 2

\qquad \qquad \qquad \qquad \qquad \qquad \qquad \qquad \qquad Fig. 3

We now turn to a theoretical description \cite%
{sorensen,sorensen2,Duan2,details}. The crucial parts of the whole setup are
the C-QED systems which supply ideal platforms to realize the strong
coupling between photons and atoms. Since the photon will be detected and
discarded, here we concentrate on the atom inside the cavity. The relevant
Hamiltonian, including the coupling of the external fields \cite{explain1},
can be expressed as ($\hslash =1$) \cite{milburn}%
\begin{eqnarray}
H &=&\omega _{0}|0\rangle \langle 0|+\omega _{1}|1\rangle \langle 1|+\omega
_{e}|e\rangle \langle e|+\omega _{h}a_{h}^{\dag }a_{h}+g(|e\rangle \langle
0|a_{h}+a_{h}^{\dag }|0\rangle \langle e|)  \notag \\
&&+i\int d\omega \kappa (\omega )[b(\omega )^{\dag }a_{h}-a_{h}^{\dag
}b(\omega )]+\int d\omega \omega b(\omega )^{\dag }b(\omega )\text{,}
\end{eqnarray}%
where $b(\omega )$ denotes the one-dimesional free-space mode and has a
standard commutation relation $\left[ b(\omega ),b^{\dag }(\omega ^{^{\prime
}})\right] =\delta (\omega -\omega ^{^{\prime }})$. Since only fields with
the carrier frequency close to $\omega _{h}$ contribute mostly to the cavity
mode, we take $\kappa (\omega )=\left( \frac{\gamma }{2\pi }\right) ^{\frac{1%
}{2}}$, where $\gamma $ is the effective cavity decay rate. To describe the
effect of the spontaneous emission, we introduce the Lindablad operator $L=%
\sqrt{\gamma _{s}}\left\vert 0\right\rangle \langle e|$, where $\gamma _{s}$
denotes the decay rate of the excited state $\left\vert e\right\rangle $ to
the ground state $\left\vert 0\right\rangle $. We also neglect the
spontaneous loss of the upper level to other states.

The Heisenberg equations of the system can be described by the
following form in rotating frame
\begin{eqnarray}
\partial _{t}\tilde{\sigma}_{11} &=&0\text{,}  \label{s11} \\
\partial _{t}\tilde{\sigma}_{00} &=&\gamma _{s}\tilde{\sigma}_{ee}-ig(\tilde{%
a}_{h}^{\dag
}\tilde{\sigma}_{0e}-\tilde{\sigma}_{e0}\tilde{a}_{h})\text{,}
\label{s00} \\
\partial _{t}\tilde{\sigma}_{10} &=&-ig\tilde{a}_{h}^{\dag }\tilde{\sigma}%
_{1e}\text{,}  \label{s10} \\
\partial _{t}\tilde{\sigma}_{1e} &=&-\frac{\gamma _{s}}{2}\tilde{\sigma}%
_{1e}-ig\tilde{\sigma}_{10}\tilde{a}_{h}\text{,}  \label{s1e} \\
\partial _{t}\tilde{\sigma}_{0e} &=&-\frac{\gamma _{s}}{2}\tilde{\sigma}%
_{0e}-ig(\tilde{\sigma}_{00}-\tilde{\sigma}_{ee})\tilde{a}_{h}\text{,}
\label{s0e} \\
\partial _{t}\tilde{\sigma}_{ee} &=&-\gamma _{s}\tilde{\sigma}_{ee}+ig(%
\tilde{a}_{h}^{\dag }\tilde{\sigma}_{0e}-\tilde{\sigma}_{e0}\tilde{a}_{h})%
\text{,}  \label{see}
\end{eqnarray}%
where we have set the detune $\Delta =\omega _{h}-\omega _{e0}=0$, $\tilde{a}%
_{h}=a_{h}e^{i\omega _{h}t}$, and $\tilde{\sigma}_{ij}=$
$|i\rangle \langle j|e^{-i\omega _{ij}t}$ with $\omega
_{ij}=\omega _{i}-\omega _{j}$. The influence of the external
fields can be taken into account by the
input-output relation%
\begin{equation}
\partial _{t}\tilde{a}_{h}=-ig\tilde{\sigma}_{0e}-\frac{\gamma }{2}\tilde{a}%
_{h}-\sqrt{\gamma }\tilde{b}_{in}(t),  \label{f1}
\end{equation}%
where $\tilde{b}_{in}(t)$ is field operator corresponding the
input flux of photons with $[\tilde{b}_{in}(t),\
\tilde{b}_{in}^{\dag }(t^{^{\prime }})]=\delta (t-t^{^{\prime
}})$. The cavity output $\tilde{b}_{out}(t)$
(Fig. 3) is connected with the input by $\tilde{b}_{out}(t)=\tilde{b}%
_{in}(t)+\sqrt{\gamma }\tilde{a}_{h}(t)$. In the above equations,
we have also omitted the terms concerning to the Langevin noise,
which has negligible contribution to the dynamics.

To see the effect of the input pulse clearly, we introduce the
Fourier
component of the operator as $\hat{A}(\omega )=\int \frac{dt}{\sqrt{2\pi }}%
\hat{A}(t)e^{i\omega t}$. From Eq. (\ref{f1}) and Eq. (\ref{s0e}),
we get
\begin{equation}
\tilde{a}_{h}(\omega )=\frac{g\tilde{\sigma}_{0e}(\omega )-i\sqrt{\gamma }%
\tilde{b}_{in}(\omega)}{\omega +i\frac{\gamma }{2}}\text{.}
\label{fields2}
\end{equation}%
\begin{equation}
\tilde{\sigma}_{0e}(\omega )=\frac{g}{\omega +i\frac{\gamma
_{s}}{2}}\int \frac{d\omega ^{^{\prime }}}{\sqrt{2\pi
}}\tilde{\sigma}_{z}(\omega -\omega ^{^{\prime
}})\tilde{a}_{h}(\omega ^{^{\prime }})\text{,}  \label{sigmaoe2}
\end{equation}%
where
$\tilde{\sigma}_{z}=\tilde{\sigma}_{00}-\tilde{\sigma}_{ee}$.

We have assumed that the transition between $|0\rangle $ and
$|e\rangle $ is
a closed optical transition, so the projector operator $\hat{P}_{z}=\tilde{%
\sigma}_{00}+\tilde{\sigma}_{ee}$ is independent of time. From Eq. (\ref{s00}%
) and Eq. (\ref{see}), we find
\begin{equation}
\tilde{\sigma}_{z}(\omega )=\frac{i\gamma _{s}}{\omega +i\gamma _{s}}\hat{P}%
_{z}|_{t=0}\delta (\omega )+\frac{2g}{\omega +i\gamma _{s}}\int \frac{%
d\omega ^{^{\prime }}}{\sqrt{2\pi }}[\tilde{a}_{h}^{\dag }(\omega
^{^{\prime }})\tilde{\sigma}_{0e}(\omega -\omega ^{^{\prime
}})-h.c]\text{.} \label{sigmaZ}
\end{equation}

In our scheme the input field is considered to be a single-photon pulse, so
the light intensity is so low that at most one photon is inside the cavity
at a time. Inserting Eq. (\ref{sigmaZ}) and Eq. (\ref{sigmaoe2}) into Eq. (%
\ref{fields2}), and omitting the terms involving more than one $\tilde{a}%
_{h}(\omega )$ operator, we obtain
\begin{equation}
\tilde{a}_{h}(\omega )=\frac{-i\sqrt{\gamma }\tilde{b}_{in}(\omega )}{\omega
+i\frac{\gamma }{2}-\frac{i\gamma _{s}\left\vert g\right\vert ^{2}\hat{P}%
_{z}|_{t=0}}{(\omega +i\gamma _{s})(\omega +i\frac{\gamma _{s}}{2})}},
\label{fields3}
\end{equation}

\begin{equation}
\tilde{b}_{out}(\omega )=\frac{\omega -i\frac{\gamma }{2}-\frac{i\gamma
_{s}\left\vert g\right\vert ^{2}\hat{P}_{z}|_{t=0}}{(\omega +i\gamma
_{s})(\omega +i\frac{\gamma _{s}}{2})}}{\omega +i\frac{\gamma }{2}-\frac{%
i\gamma _{s}\left\vert g\right\vert
^{2}\hat{P}_{z}|_{t=0}}{(\omega +i\gamma _{s})(\omega
+i\frac{\gamma _{s}}{2})}}\tilde{b}_{in}(\omega ).
\label{outfields}
\end{equation}

Now we can see clearly how the CPF gate works. When the input photon is $v$
polarized, it is reflected by the PBS and the mirror without any change. In
this situation, we get $\tilde{b}_{out}(\omega )=\tilde{b}_{in}(\omega )$.
Otherwise it transmits the PBS and enters into the cavity. For ideal input
fields, $(\gamma ,\gamma _{s})\gg \omega $, we get
\begin{equation}
\tilde{b}_{out}=\frac{-1+\frac{4\left\vert g\right\vert ^{2}}{\gamma \gamma
_{s}}\hat{P}_{z}|_{t=0}}{1+\frac{4\left\vert g\right\vert ^{2}}{\gamma
\gamma _{s}}\hat{P}_{z}|_{t=0}}\tilde{b}_{in}.  \label{outfields2}
\end{equation}%
If the atom inside the cavity is in the state $|1\rangle $, which means $%
\hat{P}_{z}|_{t=0}=0$, we have $\tilde{b}_{out}=-\tilde{b}_{in}$. However,
if $\hat{P}_{z}|_{t=0}=1$ and $\frac{\left\vert g\right\vert ^{2}}{\gamma
\gamma _{s}}\gg 1$, we obtain the approximate expression as $\tilde{b}_{out}=%
\tilde{b}_{in}$.

So far we have realized the CPF gate $e^{-i\pi |1\rangle \langle 1|\otimes
|h\rangle \langle h|}$ between a trapped-atom and a single-photon pulse.
Combining this kind of two-body interaction with proper local operations,
the desired non-local gate can be realized. In fact, general local
operations are easy to perform for photons with the help of wave-plates and
beam splitters. As to this kind of \textquotedblleft $\Lambda $%
\textquotedblright\ type atoms, single-qubit rotation \cite{Duan3,jane} can
be achieved by coupling $|1\rangle $ and $|0\rangle $ via Raman transition.
Since only the two lower levels are concerned, the decoherence caused by the
level $|e\rangle $ can be strongly suppressed.

The above-mentioned CPF gate can also be generalized to multiple-atom case.
In this situation, $\hat{P}_{z}|_{t=0}$ represents the total number of the
atom in the state $|0\rangle $. When $\frac{\left\vert g\right\vert ^{2}}{%
\gamma \gamma _{s}}\hat{P}_{z}|_{t=0}\gg 1$, we get the CPF gate between
multiple-atom and the single-photon pulse.

In order to estimate the influence of the external fields on the
atom inside the cavity, we consider the density matrix elements
explicitly. The diagonal elements are unaffected when we get a
click on the detectors (the transition $|0\rangle \longrightarrow
|e\rangle $ is optical closed). The off-diagonal
elements, however, decay due to spontaneous emission from the upper level $%
|e\rangle $ to $|0\rangle $. we can roughly estimate the effect from Eq. (%
\ref{s10}) and Eq. (\ref{s1e}).

Integrate Eq. (\ref{s1e}) and substitute the expression into Eq.
(\ref{s10})
by introducing the Fourier transformation of cavity mode $\tilde{a}%
_{h}(t)=\int \frac{d\omega }{\sqrt{2\pi }}\tilde{a}_{h}(\omega
)e^{-i\omega t}$, we find

\begin{equation}
\partial _{t}\tilde{\sigma}_{10}=-\left\vert g\right\vert
^{2}\int_{t_{0}}^{t}dt_{1}\int \frac{d\omega _{1}d\omega _{2}}{2\pi }%
e^{(i\omega _{2}-\frac{\gamma _{s}}{2})t+(\frac{\gamma
_{s}}{2}-i\omega
_{2})t_{1}}\tilde{a}_{h}^{\dag }(\omega _{1})\tilde{\sigma}_{10}(t_{1})%
\tilde{a}_{h}(\omega _{2}).
\end{equation}%
If we assume that the coherence between the two lower levels
doesn't change much on time scale $1/\gamma _{s}$, we can get the
following relation (by
setting $t_{0}\rightarrow -\infty $)%
\begin{equation}
\partial _{t}\tilde{\sigma}_{10}=-\int \frac{d\omega _{1}d\omega _{2}}{2\pi }%
\frac{\left\vert g\right\vert ^{2}}{\frac{\gamma _{s}}{2}-i\omega _{2}}%
\tilde{a}_{h}^{\dag }(\omega
_{1})\tilde{\sigma}_{10}(t)e^{i(\omega _{1}-\omega
_{2})t}\tilde{a}_{h}(\omega _{2})\text{.}
\end{equation}%
Integrating this equation we obtain the formal solution of $\tilde{\sigma}%
_{10}$ as
\begin{equation}
\tilde{\sigma}_{10}(\infty )=:\tilde{\sigma}_{10}(t_{0})e^{-\int
d\omega
\frac{\left\vert g\right\vert ^{2}}{\frac{\gamma _{s}}{2}-i\omega }\tilde{a}%
_{h}^{\dag }(\omega )\tilde{a}_{h}(\omega )}:\text{,}
\label{10end}
\end{equation}%
where $::$ denotes the normal ordering of the corresponding
operators.

Eq. (\ref{10end}) illustrates the influence of the input pulses,
and from which we can roughly estimate the quality of the
non-local gate. If we assume that the bandwidth of the input
fields is narrow and symmetrical around the resonant frequency
$\omega _{h}$, we can obtain
\begin{equation}
\tilde{\sigma}_{10}(\infty )=\tilde{\sigma}_{10}(t_{0})(1-\frac{8\frac{%
\left\vert g\right\vert ^{2}}{\gamma \gamma _{s}}}{(1+4\frac{\left\vert
g\right\vert ^{2}}{\gamma \gamma _{s}}\hat{P}_{z}|_{t=0})^{2}}).
\end{equation}%
The fidelity of the non-local gate can be approximately expressed as
\begin{equation}
F=1-2(\left\vert ab\right\vert ^{2}\Delta _{B}+\left\vert \alpha \beta
\right\vert ^{2}\Delta _{A}),  \label{fidelity}
\end{equation}%
\begin{equation}
\Delta _{A(B)}=\frac{8G_{A(B)}}{(1+4G_{A(B)}\hat{P}_{z}(t=0)_{A(B)})^{2}},
\label{delta}
\end{equation}%
where $G=\frac{\left\vert g\right\vert ^{2}}{\gamma \gamma _{s}}$, and the
subscript $A(B)$ means that the value is obtained according to the
parameters of system $A(B)$. If $\left\vert a\right\vert ^{2}$ and $%
\left\vert \alpha \right\vert ^{2}$ are large enough \cite{explain} and If
the two cavity systems $A$ and $B$ have the same parameters, we have $%
1-F\thicksim G^{-1}=\frac{\gamma \gamma _{s}}{\left\vert g\right\vert ^{2}}$%
. For current technology level, the parameter $G$ can reach 100 \cite{cavity}%
. So we can get $F\sim 0.99$ in this situation. Eq. (\ref{fidelity})
indicates that the fidelity of our scheme depends only on the magnitude of
coupling factor $g$, and not on the phase. So it is not necessary to confine
the atom within a wavelength of the field. The factor $g$ also suffers
significant random variation in current experiment due to residual atomic
motion. However, from the numerical simulation of reference \cite{Duan}, we
expect that it is also not necessary to fix the atom inside the cavity.

The major source of noise comes from the dark counts of the single-photon
detectors, which can produce false positive results. Fortunately, if there
is no photon loss, the dark counts will not affect the measurement results
of $A$ and $B$, or can be definitely distinguished when both of the two
detectors get a click on one side. This effect can be considered as a
shrinking factor $1-2p_{l}p_{dc}$ connected to the fidelity of the final
state, where $p_{l}$ is the probability of photon loss (the inefficient of
single-photon detectors can be included here as a modified photon loss rate)
and $p_{dc}$ describes the dark count rate. The input pulse may be reflected
because of the mismatch between the incident field and the cavity mode or
other imperfections. If we introduce $f$ to describe these imperfections,
the total fidelity will be reduced by $\left[ \frac{(1-f)R}{f+(1-f)R}\right]
^{2}$, where we assumed that the two cavity systems have the same reflection
coefficient $R$ (for resonant case, $R=(\frac{1-\frac{4\left\vert
g\right\vert ^{2}}{\gamma \gamma _{s}}\hat{P}_{z}|_{t=0}}{1+\frac{%
4\left\vert g\right\vert ^{2}}{\gamma \gamma _{s}}\hat{P}_{z}|_{t=0}})^{2}$%
). All these effects cause the fidelity reduced by $(1-2p_{l}p_{dc})\left[
\frac{(1-f)R}{f+(1-f)R}\right] ^{2}$. And the success probability can also
be estimated as $(1-p_{l})(1-p_{l}-2p_{dc})$.

The whole scheme can be extended to multi-party case, where more cavity
systems and entangled pairs should be contained here. If a quantum network
computation contains $N$ times such kind of C-NOT operation, the total
shrinking factor which is connected to the fidelity will be $%
(1-2Np_{l}p_{dc})\left[ \frac{(1-f)R}{f+(1-f)R}\right] ^{2N}$ (we assumed
the dark counts rate is small enough), and the corresponding success
probability is $(1-p_{l})^{N}(1-p_{l}-2p_{dc})^{N}$.

It should be pointed out that we have supposed the bandwidth of the
entangled pules is narrow compared with the linewidth of the cavity and the
atom. Such kind of entangled source can't be obtained with spontaneous
parametric down-conversion (SPDC). However, Gheri \textit{et al.}\ \cite%
{gheri} have proposed an inspiring scheme to generate entangled pairs with
the assistance of optical cavity. With this scheme the required bandwidth
could be achieved.

In summary, we have proposed a very simple scheme for future's network
quantum computation. With the C-QED system and entanglement, we have
realized the non-local gate between two different quantum network nodes.
Compared with the schemes proposed before, which often use photon as flying
qubit to transmit interaction between two separate nodes, our method
contains entangled qubit as quantum channel, so environment noise can be
suppressed. Moreover, the entanglement between the two nodes can be
implemented and stored before the whole process. For example, we can store
the entanglement with atomic ensemble \cite{ensemble}, and then by a weak
pulse, the stored entanglement can be released to complete the desired
non-local operation.

We thank M. Paternostro and A. Beige \cite{related} for drawing our
attention to their related works. This work was funded by the National
Fundamental Research Program (2001CB309300), the National Natural Science
Foundation of China (10304017) and the Innovation Funds from the Chinese
Academy of Sciences.

\bigskip

Figure captions:

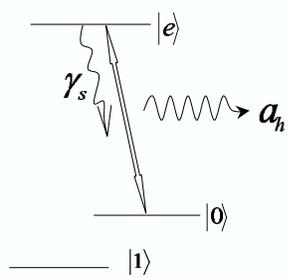
Fig. 1. Energy levels of the atom inside the cavity. An cavity
mode $a_{h}$\ resonantly drives the transition $|e\rangle
\longrightarrow |0\rangle $ with coupling strength $g$, and
$\gamma _{s}$ describes the corresponding decay rate.

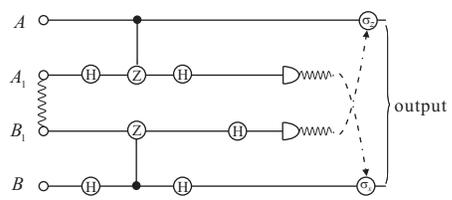
Fig. 2. Logical circuit to implement the non-local gate with $A$
as the control qubit and $B$ as the target. Classical
communication is represented by the dashed line, and suitable
local rotations can be found in Table. 1.

Fig. 3. Experimental setup of the whole scheme. The produced
entangled photon pairs enter the CPF system and reach the
detectors to cause clicks, then classical communication is
performed to complete the non-local gate. The half wave plates
(HWP) used here are to perform the local Hadamard rotation of
photons, and local operation of the atom can be realized with the
help of two classical fields.

\bigskip Table captions:

\bigskip Table. 1. Local operation for the corresponding measurement result.
For example, $D_{A_{h}}$ means the photon detected by Alice is horizontal ($%
h $) polarization.


\begin{thebibliography}{99}
\bibitem{explain1} Since the arrangement of $A$ and $B$ is symmetrical, it's
sufficient to consider only half part of the whole system.
\bibitem{explain} For small $\left\vert a\right\vert ^{2}$ and $\left\vert
\alpha \right\vert ^{2}$, the atom mostly stays in the state $|1\rangle $.
So the decay of the upper level $|e\rangle $ has little effect on the
fidelity, and we can estimate $F$ as $1-F\thicksim O(\left\vert a\right\vert
^{2}+\left\vert \alpha \right\vert ^{2})$.
\end{thebibliography}
\end{document}